\documentclass[conference]{template/IEEEtran}
\IEEEoverridecommandlockouts
\usepackage{amsmath,amssymb,amsfonts}
\usepackage{algorithmic}
\usepackage{graphicx}
\usepackage{textcomp}
\usepackage{xcolor}

\usepackage[style=ieee,mincitenames=1,maxcitenames=1,minbibnames=1,maxbibnames=2]{biblatex}
\AtBeginBibliography{\footnotesize}
\AtEveryBibitem{\clearlist{location}}

\usepackage{orcidlink}
\usepackage{cleveref} 
\usepackage{glossaries}
\usepackage{booktabs, multirow, multicol}

\usepackage{graphicx}
\glsdisablehyper

\newacronym{ai}{AI}{Artificial Intelligence}
\newacronym{tts}{TTS}{text-to-speech}
\newacronym{vc}{VC}{voice conversion}
\newacronym{sls}{SLS}{Sensitive Layer Selection}
\newacronym{ssl}{SSL}{Self-supervised Learning}
\newacronym{poi}{POI}{Person-of-Interest}

\newcommand*{\asvspoof}{ASVspoof}

\newcommand*{\sincnet}{SincNet}
\newcommand*{\rawnet}{RawNet2}
\newcommand*{\rawgat}{RawGAT-ST}
\newcommand*{\aasist}{AASIST}

\newcommand*{\xlsr}{XLS-R}
\newcommand*{\xlsraasist}{XLSR-AASIST}
\newcommand*{\xlsrsls}{XLSR-SLS}
\newcommand*{\xlsrmamba}{XLSR‑Mamba}

\newcommand*{\poi}{\gls{poi}}
\newcommand*{\ssl}{\gls{ssl}}

\makeatletter
\newcommand{\warn}[1]{\@latex@warning{#1}}

\makeatother

\usepackage{background}
\backgroundsetup{contents={\begin{tabular}{c}
     \textcopyright 2026 IEEE.  Personal use of this material is permitted. \\
     Permission from  IEEE must be obtained for all other uses, in any current or future media, including reprinting/republishing \\[-.25em]
     this material for advertising or promotional purposes, creating new collective works, for resale or redistribution to servers or \\[-.25em]
     lists, or reuse of any  copyrighted component of this work in other works.\\ 
\end{tabular}}, scale=1, placement=bottom, vshift=1.5em, angle=0, opacity=1}

\begin{document}

\title{Neural Encoding Detection is Not All You Need\\for Synthetic Speech Detection
\thanks{This paper was supported by the BMBF news-polygraph project (grant no. 03RU2U151D), and by JST, PRESTO (grant no. JPMJPR23P9).}
}

%TODO: add ORCID links?

\author{
\IEEEauthorblockN{
Luca Cuccovillo\IEEEauthorrefmark{1}\orcidlink{0000-0001-5559-6508},
Xin Wang\IEEEauthorrefmark{2}\orcidlink{0000-0001-8246-0606},
Milica Gerhardt\IEEEauthorrefmark{1}\orcidlink{0000-0003-0017-9656}, and
Patrick Aichroth\IEEEauthorrefmark{1}\orcidlink{0000-0003-4777-6335}
}\vspace{0.5em}
\IEEEauthorblockA{\IEEEauthorrefmark{1}\textit{Fraunhofer Institute for Digital Media Technology, Ilmenau, Germany}}
\IEEEauthorblockA{{\{ luca.cuccovillo, milica.gerhardt, patrick.aichroth \}@idmt.fraunhofer.de}}\vspace{0.5em}
\IEEEauthorblockA{\IEEEauthorrefmark{2}\textit{National Institute of Informatics, Tokyo, Japan}}
\IEEEauthorblockA{\{ wangxin \}@nii.ac.jp}
}

\iffalse

\author{%
\IEEEauthorblockN{Luca Cuccovillo \orcidlink{0000-0001-5559-6508}}
\IEEEauthorblockA{\textit{Fraunhofer IDMT} \\
Ilmenau, Germany\\
luca.cuccovillo@idmt.fraunhofer.de}
\and
\IEEEauthorblockN{Xin Wang \orcidlink{0000-0001-8246-0606}}
\IEEEauthorblockA{\textit{National Institute of Informatics} \\
Tokyo, Japan\\
wangxin@nii.ac.jp}
\and
\IEEEauthorblockN{Milica Gerhardt \orcidlink{0000-0003-0017-9656}}
\IEEEauthorblockA{\textit{Fraunhofer IDMT} \\
Ilmenau, Germany\\
milica.gerhardt@idmt.fraunhofer.de}
\and
\IEEEauthorblockN{Patrick Aichroth \orcidlink{0000-0003-4777-6335}}
\IEEEauthorblockA{\textit{Fraunhofer IDMT} \\
Ilmenau, Germany\\
patrick.aichroth@idmt.fraunhofer.de}
}

\fi
\maketitle
\begin{abstract}
This paper reviews the current state and emerging trends in synthetic speech detection.
It outlines the main data-driven approaches, 
discusses the advantages and drawbacks of focusing future research solely on neural encoding detection,
and offers recommendations for promising research directions.
Unlike works that introduce new detection methods or datasets, this paper aims to guide future state-of-the-art research in the field and to highlight the risk of overcommitting to approaches that may not stand the test of time.
\end{abstract}


\section{Introduction}

%Synthetic speech is the devil -- must include a *definition*
The abuse of synthetic speech, i.e., speech artificially generated by a computer or an electronic system using \gls{tts} or \gls{vc} algorithms, is a growing area of concern. After the advent of \gls{ai}, and the consequent increase in quality, synthetic speech was identified by both Europol and Interpol as a critical threat for society, with the potential to become a staple tool for organized crime and undermine the trust in digital media as an arbiter of truth in legal proceedings~\cite{europol_deepfakes, interpol_deepfakes}.

%Very fragmented landscape due to the urgency and challenge-driven nature of synthesis detection
Speech synthesis detection gained traction in the research community through the pioneering Automatic Speaker Verification Spoofing and Countermeasures (\asvspoof{}) challenges, addressing this topic since 2019~\cite{dataset_asvspoof19,dataset_asvspoof21,dataset_asvspoof5}.

The urgency of the problem, as-well-as the competitive nature of the \asvspoof{} challenge and similar initiatives such as the Audio Deep synthesis Detection (ADD)~\cite{challenge_add2022} or the Synthetic Audio Forensics Evaluation (SAFE)~\cite{challenge_safe2025} challenges, led to a fast-paced proposal of detection algorithms, producing a diverse and fragmented landscape in which data-driven approaches largely prevailed. 

%Over-reliance on data-driven == trapped by bias
Data-driven approaches brought significant advantages to the field, but did not come without important disadvantages~\cite{cuccovillo2022openchallenges}. The more complex the processing of the input data, the more abstract the embeddings used for the detection---with a striking lack of explainability that might render these models inadmissible in legal proceedings~\cite{lex_xai,lex_aiact}. Over time, data-driven models proved to strongly relate their output to silent intervals in which no speech signal is present, with a dramatic performance loss when these intervals are removed from the test content~\cite{issues__silentintervals}. Further studies used trustworthy \gls{ai} tools to show that, in addition to focusing on silent intervals, data-driven methods rely on low-frequency and high-frequency bands~\cite{issues__highfrequencies} and on background noise~\cite{issues__backgroundnoise} rather than on the characteristics of the input speech content---which indeed should be the subject matter of the detection.

%Latest emerging trend is neural encoding detection
In more technical terms, data-driven methods do not rely on traces left by the feature extraction stage of \gls{tts} and \gls{vc}, responsible for converting the input text or voice into acoustic features corresponding to the desired content and speaker identity. Rather, they rely on traces from the vocoding stage, that reconstructs the final waveform from the intermediate acoustic features---i.e., on neural encoding traces. 
The detection of neural encoding traces, that emerged as an undesired side effect of data-driven detectors, and that was proposed to temporarily mitigate the scarcity of high-quality training datasets~\cite{vocoding_detection__wang23}, is rapidly becoming the focus of research on synthetic speech detection~\cite{dataset_codecfake1,dataset_codecfake2}---ignoring the issue that such traces appear whenever acoustic features are converted into a waveform using a neural network, even when they originate from pristine natural recordings.
%These traces, indeed, are stronger than the ones left by the  feature extraction stage, and are naturally selected by data-driven approaches by the principle of least effort~\cite{issues__shotcutlearning}.

%structure of the paper
In the rest of the paper, we investigate this alarming trend in great detail. \Cref{sec:data-driven} will introduce the three main categories of existing data-driven algorithms for synthetic speech detection: SincNet-based, Self-supervised Learning (SSL)-based, and neural encoding detection.
%. \Cref{sec:neural-encoding} will define neural encoding detection and relate it to the shortcomings of existing data-driven state-of-the-art algorithms.
\Cref{sec:neural-encoding-isnt-enough}, the core of this work,  will focus on the advantages and drawbacks of narrowing down future research on neural encoding detection, and recommend research directions able to foster future progress in synthetic speech detection. \Cref{sec:hypothesis-driven} will introduce alternative hypothesis-based synthesis detection paradigms that could serve as a starting point to overcome the shortcomings of data-driven algorithms. \Cref{sec:outlook} will close with a summary of the paper and its key conclusions.

%% ============= Landscape of Synthetic Speech Detection

\section{Data-Driven Synthetic Speech Detection}\label{sec:data-driven}

This section summarizes the most influential data-driven algorithms for synthetic speech detection, the behavior of which is primarily determined by the selected training data.

% --------------------------------------
% Spectro-temporal Data-driven Methods
% --------------------------------------
\subsection{SincNet-based Methods}

%definition
Many proposals for synthetic speech detection derive from \sincnet{}~\cite{sota_sincnet}, a neural architecture designed to use raw audio samples as input, and to encode them using a set of band-pass filters.
The peculiarity of the \sincnet{} filterbank is that the width and center frequencies of each filter are expressed with pairs of cardinal sines $\operatorname{sinc}(x)=\sin(x)/x$, and can either be determined at training time via backpropagation or be fixed to reflect, e.g., a Mel scaling.

% RawNet2
The first synthetic speech detection using the \sincnet{} encoding paradigm was \rawnet{}~\cite{sota_rawnet2}, where the \sincnet{} filterbank was coupled with a series of residual blocks, the embeddings of which were decoded by gated recurrent units (GRUs) before entering the classification stage. The authors' intent was to maximize the discriminative capabilities of the model also for the most elusive attacks in the \asvspoof{}---an attempt so successful that the \rawnet{} model became a ubiquitous evaluation baseline in synthetic speech detection.

% RawGAT-ST
In order to better model the interdependencies between the time and frequency domains, in \rawgat{}~\cite{sota_rawgat} the \rawnet{} encoder was concatenated with a spectro-temporal (ST) graph attention network (GAT), which replaced the GRU-based decoder. The model leveraged two branches, one focusing on the spectral domain and the other on the temporal one, to generate features detailing different frequency bands at different time locations, overcoming the time-invariant nature of \rawnet{}.

% AASIST
Finally, in \aasist{}~\cite{sota_aasist} the spectral and temporal dimensions of the two \rawgat{} branches were merged by means of a heterogeneous stacking graph attention layer, able to correctly model the co-occurrence of artifacts between the two domains---an aspect that could not be fully exploited by the fusion solution in \rawgat{}. The performance improvement, however, came at the cost of an even higher asymmetry between the simplicity of the \sincnet{} filterbank and the complexity of the subsequent analysis stage, suggesting the presence of a possible bottleneck in the encoding stage.

%pros
The main advantage of approaches based on the \sincnet{} filterbanks, besides sheer performance, is that they avoid the overuse of 2D convolutional layers on melspectrograms---a common practice before the proposal of \rawnet{} but suboptimal since the information is highly distributed in the spectral domain, and defies the locality of convolutional filters.

%cons
On the negative side, the time-invariant nature of these filterbanks tends to concentrate the model attention on a small number of frequency bands of high significance. In particular, these models proved to converge toward high- and low-frequency regions in which speech is absent~\cite{issues__highfrequencies}, focusing on the background rather than on speech~\cite{issues__backgroundnoise}, and ultimately impairing model generalization~\cite{sota_xlsr_aasist}.

% --------------------------------------
% Self-supervised-learning-based Methods
% --------------------------------------
\subsection{\texorpdfstring{\ssl{}}{Self-supervised Learning (SSL)}-based Methods}

%definition
Most \ssl{}-based algorithms rely on an encoding backbone employing a wav2vec~2.0 model pretrained to generate cross-lingual speech representations (\xlsr{})~\cite{baevski2020wav2vec,babu2022xls}.

The wav2vec~2.0 model operates on the time domain, by combining a series of 1D convolutional layers---realizing a filterbank, even though not explicit as in \sincnet{}---with a transformer model. The output embeddings of the transformer, that capture correlations between different filters---hence, frequency bands---are trained to identify the true quantized latent speech representation for a masked time step within a set of distractors. In the \xlsr{} variant, this self-supervision mechanism was performed across 128 languages and relied on more than two billion parameters. %a much higher number of parameters (2 billion) than the ones in the initial wav2vec~2.0 (0.3 billion). 

% XLSR-AASIST
Due to the \ssl{} mechanism, models relying on the \xlsr{} encoder have an intrinsically higher generalization capability than the ones based on the \sincnet{} filterbank. 
The first model of this kind  was \xlsraasist{}~\cite{sota_xlsr_aasist}, in which the \aasist{} architecture was revised by replacing the \sincnet{} filterbanks with the \xlsr{} embeddings, leaving the downstream \rawnet{} encoder and \aasist{} graph-based layers intact.
The use of the \xlsr{} embeddings coupled with the \aasist{} graph-attention post-processing led to better performance and noticeably lower generalization issues, that were entirely attributed to the revised encoding.

% XLSR-SLS
In an attempt to fully exploit the attention mechanism inherent in the wav2vec~2.0 model, \xlsrsls{}~\cite{sota_xlsr_sls} proposed to revise the \aasist{} decoding by gathering information not only from the very last output block of the \xlsr{} architecture, but rather from all hidden embeddings of each transformer layer, assuming that insightful information for synthetic speech detection could be retrieved at different depths---and hence, abstraction levels---of the transformer network. 
Hence, within \xlsrsls{} the information is condensed at different depths by means of a layer-wise weighted sum, with weights determined by a novel module for \gls{sls}. The \gls{sls} module attributes larger weights to the layers with the highest temporal dynamics. The weighted sum was then reduced by a series of max-pooling, fully-connected, and softmax layers to output the final decision.

% XLSR-Mamba
Most recently, in \xlsrmamba{}~\cite{sota_xlsr_mamba} the \aasist{} graph-attention-based post-processing was replaced by a bi-directional Mamba state space model. 
Mamba is a sequence-modeling architecture alternative to attention-based transformers that does not compute pairwise interactions between all input tokens, but that rather processes inputs sequentially and models their dynamics by means of its internal space. In \xlsrmamba{}, the input sequence is processed once in the forward and once in the backward time direction, and then used to predict the output label---like a class token of a standard transformer.
Compared to other \ssl{}-based models, \xlsrmamba{} captures long-range feature information more efficiently, and is able to correctly model both the  short-term anomalies and the long-term correlations, despite the reduced number of parameters.

%pros
Models based on an encoding stage based on \ssl{}, such as the \xlsr{} model, can profit from a rich input embedding stage and constitute the current state of the art in synthetic speech detection. The lack of explicit weaknesses in the model design suggests that potential limitations in performance or generalization are primarily due to the selection of training data, underscoring the critical importance of its quality.

%cons 
The main weakness of existing \ssl{}-based models resides in their lack of explainability, since current post-hoc explainability methods for images fall short in determining which characteristics of the input audio are the most relevant for the synthesis detection task. This issue is of paramount importance, since it might render them inapplicable in legal proceedings~\cite{lex_xai,lex_aiact}.

% Definition of neural encoding detection
%\section{Neural Encoding Detection}\label{sec:neural-encoding}
\subsection{Neural-encoding-detection Methods}

% self-vocoding for dataset generation
A seminal contribution by \citeauthor{vocoding_detection__wang23}~\cite{vocoding_detection__wang23} proposed an innovative approach to dataset generation for training synthetic speech detectors. The method involved extracting acoustic features from natural speech recordings, synthesizing waveforms using multiple neural vocoders, and using the resulting audio as examples of ``synthetic'' speech for training purposes. This self-vocoding pipeline enabled the construction of large-scale training datasets without requiring access to diverse TTS systems. The experiments demonstrated that  \ssl{}-based models are particularly effective at capturing neural encoding traces introduced by vocoders.

% self-vocoding for synthetic speech detection
Building upon this idea, \citeauthor{vocoding_detection__sun23}~\cite{vocoding_detection__sun23} introduced a multi-task architecture for simultaneous vocoder identification and synthesis detection. Their model incorporated a \sincnet{} filterbank as a front-end feature extractor, specifically tailored to capture the spectral patterns left by neural vocoders. This study confirmed that \sincnet{}-based models can effectively learn vocoder-specific artifacts. However, it also revealed a critical limitation: the learned detectors often failed to generalize. Audio re-synthesized using vocoders not included in the training set was rarely identified as synthetic, highlighting the strong overfitting tendencies of such systems.% and emphasizing the need for improved generalization techniques.

% self-vocoding detection is not synthetic speech detection
Despite their technical merits, these studies \emph{did not} address synthetic speech detection in the strict sense. In standard definitions, synthetic speech refers to audio generated by \gls{tts} or \gls{vc} systems, where intermediate acoustic features are derived from input text or produced by modifying an input utterance. In contrast, the aforementioned studies used natural speech as input and applied vocoding operations, meaning that the core acoustic features remained natural. As such, these works should be classified under the emerging research area of neural encoding detection---the task of identifying whether audio has undergone processing by a neural codec or vocoder, without necessarily implying that the audio was artificially synthesized from scratch.
%% One slightly improved pipeline\cite{vocoding_lu_improving_2024} intentionally degraded the acoustic features extracted from the natural waveform before they are fed to the vocoders. However, the degradation is not equivalent to the conversion done by a TTS and VC acoustic model: The performance improvement is limited, and the issue remains.

% explicit definition and characteristics of neural audio codecs detection
The task of neural encoding detection was further clarified and formalized by \citeauthor{vocoding_detection__moussa24}~\cite{vocoding_detection__moussa24}, who systematically investigated the detectability of artifacts introduced by commercially relevant neural audio codecs, such as \copyright{}Google's Lyra-V2, \copyright{}Meta's EnCodec, and Descript Inc.'s Audio Codec. These codecs, likely to become standard components in future speech synthesis pipelines, were shown to introduce strong and consistent artifacts in the high-frequency domain. These traces were so prominent that a simple logistic regression applied to the high-pass-filtered Fourier transform of the input could detect them with performance comparable to deep learning models---at least on in-domain data. On out-of-domain data, however, both simple logistic regression and complex deep learning models failed similarly: Neural encoding detection turned out to be surprisingly easy, but hardly generalizable.

\begin{figure}[ht]
\centerline{\includegraphics[width=\columnwidth]{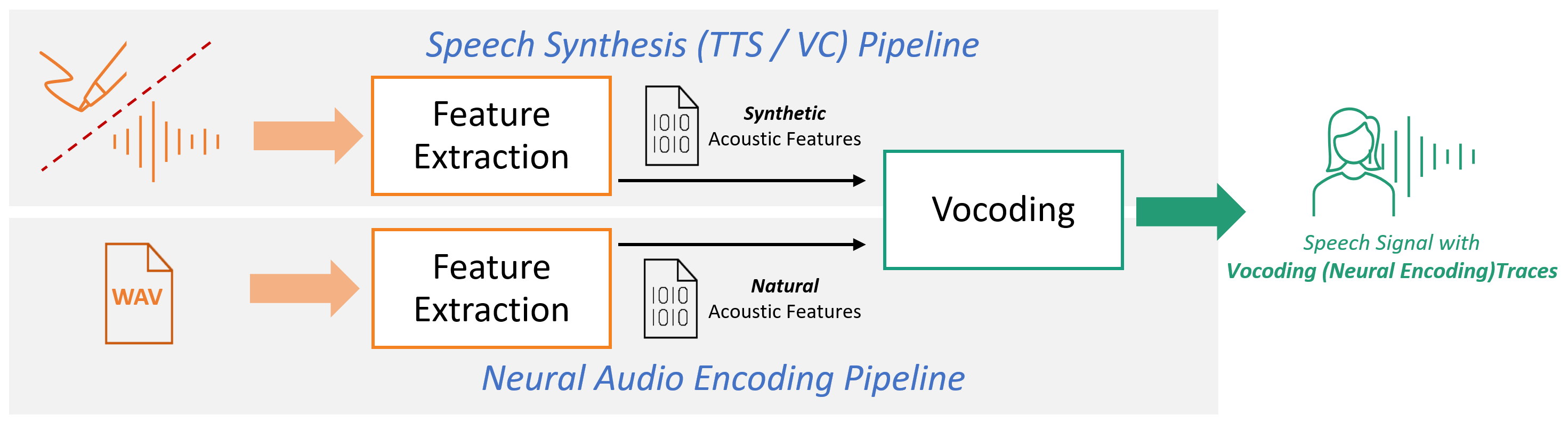}}
\caption{Speech Synthesis and Neural Audio Encoding Pipeline}
\label{fig:encoding-pipeline}
\end{figure}

% Purely data-driven algorithms are neural encoding detectors
The work by \citeauthor{vocoding_detection__moussa24}~\cite{vocoding_detection__moussa24} helped underline an important point: Neural encoding detection is essentially vocoding trace detection. If we accept this equivalence, it becomes clear that most data-driven methods for synthetic speech detection have actually been responding to encoding artifacts introduced by the vocoder stage---rather than detecting anomalies in the acoustic features themselves. This interpretation, also supported by the striking commonalities between synthesis and neural encoding pipelines in \Cref{fig:encoding-pipeline}, helps explaining several well-known shortcomings of neural detectors: their focus on silent intervals~\cite{issues__silentintervals}, sensitivity to frequency bands irrelevant for speech~\cite{issues__highfrequencies}, and reactions to background noise~\cite{issues__backgroundnoise}. These common shortcomings arise from the fact that data-driven models tend to respond not to the speech being generated or natural, but to the signal being neurally encoded or not: Since neural encoding traces are often more salient than the subtle artifacts of synthesis, models are drawn to them by the principle of shortcut learning~\cite{issues__shotcutlearning}.

\section{Is Neural Encoding Detection Enough?}\label{sec:neural-encoding-isnt-enough}

%point out the issue

In the short term, relying on neural encoding traces---i.e., on unconstrained data-driven approaches---for synthetic speech detection may appear appealing: the number of used vocoders is relatively limited, and collecting reference data for encoding artifacts is less demanding than replicating full-synthesis pipelines; it may also help uncover shared characteristics among vocoder families using similar waveform generation paradigms, such as diffusion-based or GAN-based models. 

In the long term, however, this focus poses significant risks. As long as neural encoding artifacts remain exclusive to synthesized audio, such models will continue to perform well on \acrlong{tts} and \acrlong{vc} detection tasks. But once neural codecs become the norm and replace traditional lossy encoders, these artifacts will no longer constitute a reliable signal of synthesis, rendering existing data-driven solutions suddenly obsolete and unreliable.

\begin{figure*}[t]
\centering
\includegraphics[width=0.5\linewidth]{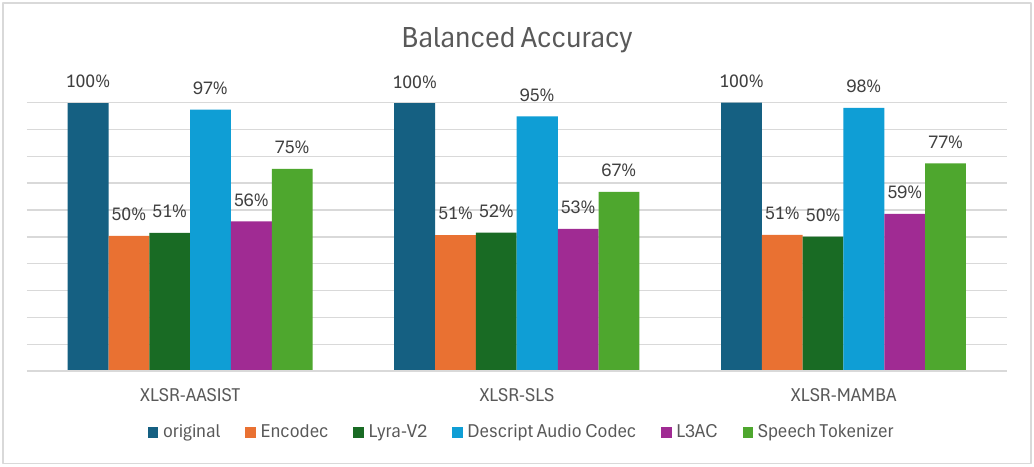}%
\includegraphics[width=0.5\linewidth]{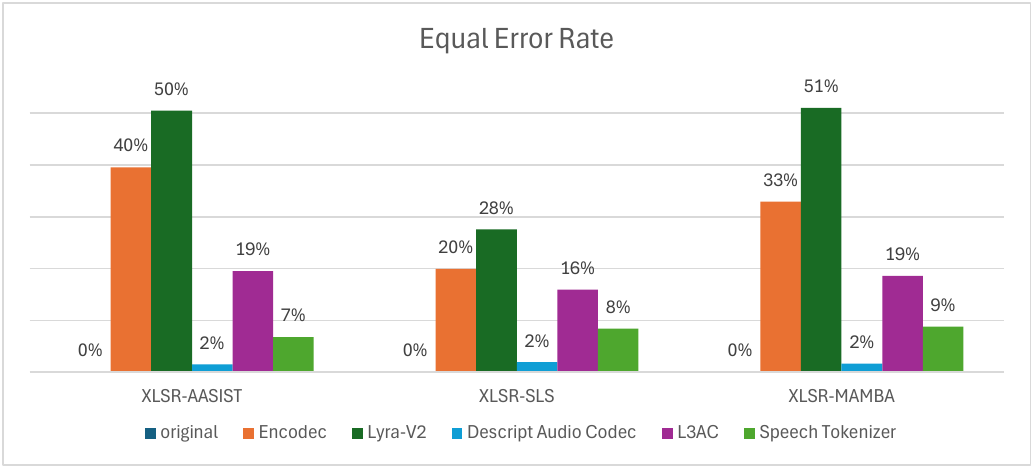}%
\caption[Performance on the eval partition of ASVspoof19 LA]{Performance of SSL-based models on the ASVspoof19 LA eval dataset~\cite{dataset_asvspoof19}, with varying neural encoders applied to the bona fide trials.}
\label{fig:experiments}
\end{figure*}

The impact of this shortcoming is evident in \Cref{fig:experiments}, where we compare the balanced accuracy (BAC) and equal error rates (EER) for the most recent SSL-based methods---using the weights provided by their authors---on the pristine \asvspoof{} 2019 LA eval dataset with those obtained on variants created by neurally encoding the bona fide trials. With the exception of the Descript Audio Codec, detection performance degrades significantly in presence of neural encoding, highlighting the need for urgent action and countermeasures \footnote{Visit \url{https://neural-isnt-deepfake.github.io} to access the neurally encoded bona fide trials and fully detailed results, including alternative test conditions supporting the finding: re-encoded spoof trials with unmodified bona fide, and fully re-encoded eval dataset.}.

\warn{TODO: update footnote with url to a github.io website}

%future direction #1 -> datasets
To address this problem in future data-driven approaches, neural encoding of natural data---using the same vocoders as those applied to synthetic data, while preserving the ground-truth label of the resulting signals as \emph{natural}---should become standard practice. The absence of obvious shortcuts may encourage existing models to focus on anomalies created by feature extraction, and thus on synthetic speech detection. Furthermore, it could lessen the security issue inherent in purely data-driven models: The possibility to perform low-cost false-positive attacks via  neural encoding.

Implementing such a solution, however, requires collaborative efforts and financial investments. Many existing datasets suffer from insufficient documentation, inadequate consent procedures, or limited speaker diversity, and are often developed by institutions focused on synthesis detection. To alleviate the persistent scarcity of high-quality synthetic examples, collaboration with companies and research institutions developing state-of-the-art \gls{tts} and \gls{vc} algorithms is essential.

%future direction #2 -> standard benchmarking

Likewise, benchmarking of synthetic speech detection solutions should be standardized upon datasets with clear characteristics: Algorithms for synthetic speech detection, ideally, should reach perfect scores on, e.g., the \asvspoof{} datasets~\cite{dataset_asvspoof19,dataset_asvspoof21,dataset_asvspoof5}, but systematically output ``natural'' for test files drawn from, e.g., the CodecFake datasets composed of entirely self-vocoded material~\cite{dataset_codecfake1,dataset_codecfake2}.

%future direction #3 -> post-hoc explainability
Post-hoc explainability approaches for audio signals should be proposed and applied to existing and future synthetic speech detection algorithms. Even though the network design might reduce the risk of shortcut learning, unwanted sample bias might still affect the results in unpredictable ways. These approaches have the potential to identify anomalous behaviors, thereby highlighting anomalies in the training distributions.

%future direction #N -> ex-ante explainability

The natural tendency of neural networks toward shortcut learning can be addressed by leveraging expert models specifically designed with explainable detection mechanisms~\cite{ghosh2023icmlw}. These so-called hypothesis-based approaches deserve further exploration, as they promote explainable-by-design architectures that are inherently more resilient to shortcut learning. Moreover, their transparent, non–black-box nature may make them suitable for legal proceedings. Following this rationale, the next section presents several such algorithms, which may serve as a basis for alternative approaches or future research in the same direction.

% Moved from above -- becomes an extended "ideas for future work" section

\section{\texorpdfstring{Hypothesis-Driven\\Synthetic Speech Detection}{Hypothesis-Driven Synthetic Speech Detection}}\label{sec:hypothesis-driven}

This section reports a few examples of hypothesis-driven algorithms for synthetic speech detection, the behavior of which is constrained by an ex-ante hypothesis on the characteristics and shortcomings of synthetic speech.

%Prosody-based Methods
\subsection{Prosody-based Methods}

%definition
Prosody-based algorithms are designed to rely on prosody-related features, i.e., on features describing the rhythm, stress, and intonation patterns of speech, and conveying meaning, emotion, and intent beyond the words themselves~\cite{hirst2024prosody}.

% joint speaker + prosody analysis
One early synthetic speech detection following this line of research proposed to combine speaker embeddings describing the timbre in the input speech, with prosody embeddings detailing its variations in rhythm, pitch and accents~\cite{prosody__attorresi2022}. The hypothesis of the work was that modeling both physiological and behavioral characteristics would have led to a semantically rich representation, exhibiting anomalies in synthetic signals.
% Furthermore, since both embeddings were robust towards lossy encoding, the method could inherently handle low-quality files from social media.

% emotion embeddings
Prosodic patterns are especially important for their capacity to convey emotions, beyond the literal verbal content. Therefore, \citeauthor{emotions__hosler2021}~\cite{emotions__hosler2021} proposed  a synthetic speech detection method based on continuous emotion descriptors. Emotional states were described by arousal (calm to agitated) and valence (negative to positive feeling) soft indicators, which capture the emotional profile of the input utterance. Conversely, \citeauthor{emotions__conti2022}~\cite{emotions__conti2022} proposed to rely on discrete emotional categories such as happiness, sadness, and anger. Despite methodological differences, both studies are grounded in the same hypothesis: that synthetic speech often exhibits imperfections in emotional expressivity, which can be exploited for detection.

% formant analysis
Since the melody of the voice is an essential element of prosody, two independent studies proposed to use a lower semantic level, and to rely on jitter and shimmer of the fundamental frequency $F_0$ for synthetic speech detection, assuming that anomalies in the pitch---and hence in $F_0$---only occur in artificial content~\cite{prosody_li10207023, prosody_unoki_deepfake_2025}. The existence of anomalies in the formant distribution of synthetic speech was also the main hypothesis at the basis of the 
speech formant analysis transformer by \citeauthor{formants__cuccovillo2024_cvpr}~\cite{formants__cuccovillo2024_cvpr}, in which $F_0$ was modeled jointly with the formants $F_1$ and $F_2$.
%Speech Formant Analysis Transformer network (SFAT-Net)~\cite{formants__cuccovillo2024_cvpr}%: In this multitask architecture, a first head is responsible for autoencoding the input speech---describing all harmonics of the speech---a second for jointly tracking both $F_0$ and the formants $F_1$ and $F_2$---describing which vowels are pronounced how---and a third one for predicting the input utterance being synthetic or natural by means of the resulting feature bottleneck.

% pros
The main advantage of prosody-based algorithms lies in their inherent explainability-by-design. Unlike purely data-driven methods---whose decisions often rely on features that defy explanation---these models operate under stronger constraints, making their behavior potentially more transparent and less susceptible to unpredictable biases.

% cons
On the negative side, prosody-based approaches are by their own nature language dependent. Features considered normal in one language---such as nasal vowels in French---may appear anomalous in another. Similarly, intonation carries lexical meaning in some languages---e.g., Mandarin Chinese---but serves an expressive function in others. Even certain sound types, like pharyngeal or ``guttural'' consonants, may be common in some languages but absent in others, complicating even further cross-linguistic generalization of these algorithms.

%POI-based Methods
\subsection{\texorpdfstring{\poi{}}{Person-of-Interest (POI)}--based Methods}

% definition
\poi{}-based algorithms reframe synthetic speech detection into verifying whether a speech segment is consistent with the claimed speaker's identity or not.
The rationale behind these approaches is that the generalization issues of data-driven models, often caused by biases in training samples, can be mitigated by modeling the natural speech of a single specific individual using a set of reference recordings. This eliminates the need to rely on a fixed set of synthesis methods and thus avoids introducing implicit sample bias.

% v1: speaker recognition embeddings
Initial approaches for \poi{}-based synthetic speech detection~\cite{poi__chen2021_xvectors,poi__pianese2022} proposed to characterize the reference recordings by means of a reference set of speaker embeddings, and then reject as synthetic all recordings for which the maximum embedding similarity to the reference set was lower than a predetermined threshold. Therefore, these works re-interpreted the problem as a speaker verification one, assuming that synthetic voices---even if defying the ears---were nevertheless unable to replicate the speaker identities accurately enough to withstand scrutiny via automatic speaker verification tools.
%
% v2: audio-LLM embeddings
A follow-up study employed large pretrained models to extract reference embeddings for \poi{}-based synthetic speech detection~\cite{poi__pianese2024}, with the goal of minimizing sample bias via the extensive training data available in audio large language models.
%Among the several alternative architectures investigated by the authors, the one achieving the best results was the BEATs transformer for latent representation learning, optimized by \acrlong{ssl} for arbitrary audio tasks~\cite{chen2023beats}, rather than cross-language speech representation as the \xlsr{} model.

% v3: interpretable speaker embeddings
The use of large pretrained models has led to improved performance in \poi{}-based synthetic speech detection, but at the cost of interpretability: These models provide no guarantee that the generated embeddings reflect the characteristics of the reference speaker, as they may instead reflect the acoustic or channel conditions present in the reference audio. Moreover, they typically lack mechanisms to indicate which aspects of the input speech matter the most for the similarity comparison. To address these limitations, \citeauthor{poi__pianese2025}~\cite{poi__pianese2025} introduced a formant-conditioned network that generates speaker embeddings based on the distribution of speech formants in the input signal. Similarly, \citeauthor{poi__salvi2025_phonemes}~\cite{poi__salvi2025_phonemes} proposed constructing speaker profiles through statistical analysis of the phonemes detected in the input recordings.

%pros
\poi{}-based models combine the advantages of a clear detection mechanism, with those deriving from one-class learning. Their output depends on a comparison with a specific reference set, making the detection rationale clear and verifiable. At the same time, they are not dependent on an arbitrary selection of known synthesis algorithms, making them generalizable and better suited to unseen attack types than data-driven methods. Additionally, they may profit from advances in speech and speaker modeling potentially leveraging richer and more interpretable feature representations over time.

%cons
On the negative side, the effectiveness of these models heavily depends on the quality and representativeness of the reference data. If the reference recordings are outdated---e.g., reflecting a younger voice---or captured under different acoustic conditions, the risk of false alarms increases significantly. Performance might be improved by coupling the positive references of the \gls{poi} with negative ones from other similar speakers or synthetic utterances, but the correct retrieval of appropriate data from the reference pool has proven to be challenging~\cite{poi__kang2024_referenceretrieval}.
%Channel mismatch due to reverberation, compression, or other distortions can also impair performance. 
Also challenging is the choice of an appropriate detection threshold, often depending on the quantity and quality of the reference set. While some work has explored self-calibration~\cite{poi__leroux2025}, a broader statistical understanding of the output behavior of these systems is still lacking.

\section{Conclusions and Outlook}\label{sec:outlook}

Despite the ongoing efforts, synthetic speech detection remains an open challenge: The performance of synthetic speech detection systems is not increasing at an acceptable pace, and as synthesis methods grow more diverse and sophisticated, existing algorithms become obsolete. 

In this paper we tried to orient future research in the domain by focusing our discussion on the distinction between neural encoding detection and synthetic speech detection: the first task addresses the detection of neural encoding artifacts produced by the vocoding stage of \gls{tts} and \gls{vc} pipelines; the second task addresses the detection of anomalies generated by the acoustic feature extraction stage.

Even if in the short term the two tasks coincide, in the long term the neural encoding detection will likely be unable to stand the test of time, and will need to be superseded by a novel generation of synthetic speech detection algorithms. Therefore, we outlined potential research directions that might be pursued in the future: the creation of collaborative, open datasets of synthetic speech; the proposal of standard benchmarking procedures and metrics; the development of post-hoc explainability methods able to identify anomalous behavior in the detection decision; the design of hypothesis-based algorithms addressing the detection in an interpretable way.

Beyond the synthetic artifacts detection framework discussed in this paper, proactive methods such as speech watermarking~\cite{cao2025watermarking} may be applicable in certain scenarios: If speech synthesis systems can automatically embed a watermark into their outputs, the problem of synthetic speech detection can be cast into a watermark detection task.
%However, before the robustness of watermark is fully solved~\cite{reilly2025deepwatermark}, we still need synthetic artifacts detection.
However, until watermarking techniques achieve sufficient robustness~\cite{reilly2025deepwatermark}, and given the presence of legacy systems and malicious actors that will not embed recognizable watermarks, passive detection methods will remain indispensable and play a primary role.

\printbibliography

\end{document}